# ISOMORPHISM BETWEEN GALILEAN AND LORENTZ TRANSFORMATIONS AND SPECIAL RELATIVITY IN GALILEAN ALGEBRA

**Mushfiq Ahmad**
Department of Physics, Rajshahi University, Rajshahi, Bangladesh
E-mail: mushfiqahmad@ru.ac.bd

**Abstract**

We have shown that Lorentz transformation is not necessary to present relativistic relative velocity. A modified definition of velocity together with Galilean transformation is equivalent to relativistic relative velocity. On the other hand, Galilean relative velocity can be presented as a Lorentz transformation. Again the definition of velocity requires a modification.

## 1. Introduction

Galilean and relativistic kinematics are distinguished by the corresponding transformation – Galilean and Lorentz. Galilean and Lorentz transformation are related by isomorphic transformations. It is, therefore, possible to represent Galilean physics with Lorentz transformation algebra and vice versa. The two physical situations, Galilean and Lorentz, will, then differ by the way velocity is defined. We shall try to show this in this paper.

## 2. Isomorphic Transformations

We shall consider motion in one space dimension.

Consider the transformation relation between primed and unprimed quantities

$$a = \frac{c}{2} \ln \frac{c+a'}{c-a'} \quad (2.1)$$

and

$$a' = c \frac{\exp\left(\frac{2a}{c}\right) - 1}{1 + \exp\left(\frac{2a}{c}\right)} \quad (2.2)$$

so that

$$a - b = d = \frac{c}{2} \ln \frac{c+d'}{c-d'} \quad (2.3)$$

where

$$d' = a'{}_c \oplus_0 (-b') = \frac{a' - b'}{1 - \frac{a'b'}{c'^2}} \quad (2.4)$$

Galilean and Relativistic Kinematics.

If $a$ is the Galilean velocity of a body and $b$ is velocity of the observer, $d$ (2.3) gives the relative velocity in Galilean representation, while $d'$ (2.4) describes the same relative velocity in Lorentz representation. This is a Galilean situation.

If $a'$ is the relativistic velocity of a body and $b'$ is (relativistic) velocity of the observer, $d'$ (2.4) gives the relative velocity in relativistic representation while, $d$ (2.3) describes the same motion in Galilean representation. This is a relativistic situation.

The same relations (2.3) and (2.4) can describe both a Galilean and a relativistic situation. Whether we are actually in a Galilean or a relativistic situation depends upon which of the two sets of quantities ($a$, $b$) or ($a'$, $b'$) is taken as the measured velocities i.e. how we define velocity.

Definition of Velocity.

Let $x$ be the distance traveled in time $t$. We shall consider 3 definitions o velocity

Definition 1.
$$a = x/t \tag{2.5}$$

Definition 2. [1]
$$a'/c = (T/t) \otimes_0 (x/cT) = \frac{(1+x/cT)^{T/t} - (1-x/cT)^{T/t}}{(1+x/cT)^{T/t} + (1-x/cT)^{T/t}} \tag{2.6}$$

$$a' \xrightarrow{T \to \infty} c \frac{\exp\left(\frac{2x}{ct}\right) - 1}{\exp\left(\frac{2x}{ct}\right) + 1} \tag{2.7}$$

$$a' \xrightarrow{x/ct \to \infty} c \tag{2.8}$$

Definition 3.
$$a''/c = \left(\frac{T}{t}\right) \overline{\otimes} \left(\frac{x}{cT}\right) = \frac{1}{2} \ln \frac{1 + \frac{T}{t} \frac{\exp\left(\frac{2x}{cT}\right) - 1}{\exp\left(\frac{2x}{cT}\right) + 1}}{1 - \frac{T}{t} \frac{\exp\left(\frac{2x}{cT}\right) - 1}{\exp\left(\frac{2x}{cT}\right) + 1}} \tag{2.9}$$

$$a'' \xrightarrow[T \to \infty]{} \frac{c}{2} \ln \frac{1 + \frac{x}{ct}}{1 - \frac{x}{ct}} \tag{2.10}$$

$$a'' \xrightarrow[x \to ct]{} \infty \tag{2.11}$$

Galilean Kinematics

If we define velocity as in Definition 1, the relative velocity will be given by
$$a - b = d \tag{2.12}$$
If we define velocity as in Definition 2, the relative velocity will be given by
$$d' = \frac{a' - b'}{1 - \frac{a'b'}{c'^2}} \tag{2.13}$$

Relativistic Kinematics

If we define velocity as in Definition 1, the relative velocity will be given by
$$d = \frac{a - b}{1 - \frac{ab}{c^2}} \tag{2.14}$$
If we define velocity as in Definition 3, the relative velocity will be given by
$$d'' = a'' - b'' \tag{2.15}$$

### 3. Galilean and Relativistic Definitions of Velocity

We know that Galilean composition of velocities (2.12) gives Galilean relative velocity (2.12). (2.12) also gives relativistic relative velocity (2.15), if the velocity is defined as in Definition 3.

We know that relativistic composition of velocities (2.14) gives relativistic relative velocity. (2.14) also Galilean relative velocity (2.13) if the velocity is defined as in Definition 2

Therefore, Galilean and relativistic kinematics may be distinguished by the way velocity is defined, keeping the same law of composition of velocities, just as they can be distinguished by the law of composition, keeping the same definition of velocity.

### 4. Einstein's Postulate

If we represent relativistic kinematics in Galilean algebra, Einstein's postulate still remains valid, velocity is redefined in such a way that the velocity becomes ∞ when $x \to ct$.

## 5. Space Contraction and Time Dilation

Space contraction and time dilation are consequences of Lorentz transformation. We have seen in section 3 that Galilean kinematics can also be presented in Lorentz transformation algebra, and conversely relativistic kinematics may be presented in Galilean algebra. This requires a re-examination of space contraction and time dilation phenomena.

## 6. Conclusion

We have seen that an alternative way of looking at Einstein's postulate and special relativity is to redefine velocity. It remains to be seen what happens to space contraction and time dilation.